\documentclass[10pt,preprint]{aastex}
\usepackage{amsmath}

\newcommand{\inctab}[1]{\includegraphics[scale=.5]{#1}}
\newcommand{\inctabp}[1]{\includegraphics[scale=.7]{#1}}
\begin{document}
\title{Topology of Neutral Hydrogen Within the Small Magellanic Cloud}

\author{ A. Chepurnov\altaffilmark{1}, A. Lazarian\altaffilmark{1}, 
J. Gordon\altaffilmark{1,2}, and S. Stanimirovic\altaffilmark{1}}

\altaffiltext{1}{Astronomy Department, University of Wisconsin,
    Madison, WI 53711}
\altaffiltext{2}{Physics \& Astronomy Department, University of Georgia,
    Athens, GA 30605}

\begin{abstract} In this paper, genus statistics have been applied to an HI column density map of the Small Magellanic Cloud in order to study its topology. To learn how topology changes with the scale of the system, we provide the study of topology for column density maps at varying resolution. To evaluate the statistical error of the genus we randomly reassign the phases of the Fourier modes while keeping the amplitudes. We find, that at the smallest scales studied ($40 \mbox{ pc}\leq\lambda\leq 80
\mbox{ pc}$) the genus shift is in all regions negative, implying a clump topology. At the larger scales ($110 \mbox{ pc}\leq\lambda\leq 250 \mbox{ pc}$) the topology shift is detected to be negative in 4 cases and positive (``swiss
cheese'' topology) in 2 cases. In 4 regions there is no statistically significant topology shift at large scales.
\end{abstract}


\section{Introduction}
The Small Magellanic Cloud (SMC) is an irregular dwarf galaxy\footnote[3]{We assume a distance of 60kpc to the SMC for all calculations throughout this paper.}. Although the SMC is believed to be gravitationally bound to the Milky Way, recent studies suggest that this may not be true \citep{B07}. The SMC is relatively metal poor and very gas-rich, and represents an ideal galaxy to study the formation of stars in low-metalicity environments. Its proximity to the Large Magellanic Cloud (LMC) and Milky Way has resulted in a turbulent history. Recent evidence points to a close encounter with the LMC some 1.5 GYr ago which deformed the SMC, creating a long, thin filament of HI called the Magellanic Stream \citep{M74,MF,S02}. The interaction between the SMC and LMC is still dynamic, with observations showing that metal poor star clusters with ages ~$<$ 200 MYr and a metalicity ratio [Fe/H] $<$ -0.6 within the LMC originated from the infalling SMC gas \citep{BK07}.

Statistical techniques are indispensable while studying turbulence in the interstellar medium (ISM). Images of neutral hydrogen (HI) distribution are frequently used as in most cases it is possible to ignore self-absorption \citep{L95}. HI also occupies a large portion of the Galactic disc (roughly a 20\% filling factor) and its movements should reflect large-scale turbulence \citep{L99}. Furthermore, the prevalence of HI means that it can be studied not only in our galaxy but in
nearby galaxies as well. 

Several studies have performed statistical analyses of the HI distribution in the SMC \citep{S99,S02}. These studies determined the spatial power spectrum of the HI intensity and employed the velocity channel analysis (VCA) technique (Lazarian \& Pogosyan 2000) to reveal shallower-than-Kolmogorov velocity and density spectra. However, power spectra, while being informative about the distribution of the energy with scale, are not sensitive to gas topology. For instance, HI surveys of the SMC have shown numerous filamentary structures and shells of expanding gas. These surveys have detected 501 shells dispersed throughout the SMC (see Fig. \ref{Shells}), six of which have radii $>$ 350pc and are large enough to be classified as supergiant shells (SGSs) \citep{SS97}. Many of these shells were created by massive associations of OB stars, but some show no spatial correlation to any young stellar population. These `orphan' shells may be a result of gamma ray bursts or collisions between high velocity clouds and the SMC \citep{H05}. Alternatively, they could be produced by the ISM turbulence.

This calls for employing other techniques for the statistical studies of SMC. Recent years have been marked by an increased interest to statistical studies of astrophysical turbulence (see Lazarian 2004, for a review). In particular, two techniques Velocity Channel Analysis (VCA) and Velocity Coordinate Spectrum (VCS) which are capable of studying  power spectra of velocity and density have been developed (Lazarian \& Pogosyan 2000, 2004, 2006). However, the topology of the gas distribution cannot be described by the power spectrum. On the contrary, genus analysis, with its ability to accurately describe and quantify the topology, is a promising synergetic tool in the quest to better understand turbulence.

Genus statistics was developed to study the topology of the universe and distribution of galaxies in three \citep{GMD86,GWM87} and two (Coles, 1986, 1991,  \citep{M89},  Plionis et al. 1992, Davis \& Coles, 1993, Coles et al. 1993) dimensions.  Subsequent projects have used genus statistics to study the topology of the temperature variations within the  Cosmic Microwave Background \citep{S94,CG03} and have also  been applied to a systematic study of the variations of  the two-dimensional genus with MHD simulations \citep{KLB07}.  The use of genus statistics for the study of HI was first discussed in \cite{L99}, and subsequent studies presented the first  genus curves for the SMC \citep{L02,L04}.  A recent paper by Kim \& Park  (2007) provided a more thorough study of the topology of  the HI peak brightness temperature distribution in the LMC. 

The goal of the present paper is to extend the  technique in Lazarian (2004) and Kim \& Park  (2007) by providing a quantitative measure of the uncertainties involved in the genus studies. We then apply the technique on the SMC HI data set to quantitatively describe the topology of the HI distribution. Kowal  et al. (2007) have shown that genus statistics of a 3D density distribution
from synthetic observations obtained with MHD simulations agrees with the genus results for the 2D integrated column density of the same data set. Therefore, genus statistics complements the power spectrum analyses in providing insights into the physical processes that shape the ISM. 

The structure of this paper is organized as follows. \S 2 discusses our approach to the analysis of data. \S 3 summarizes the
observational procedure and subsequent data analysis of the HI column density map. Section 4 is an analysis of the cropped regions within the SMC, focusing on the genus shift and its topological implications. Section 5 discusses the results and posits astrophysical connections between the genus analysis and the SMC. 

In cosmological studies the distributions in question, e.g. the distribution of CMB intensity are nearly Gaussian and genus is used to study small deviations from the Gaussian. Dealing with the ISM, in particularly, as in this paper, with the distribution of column densities of SMC, one cannot expect deviations from symmetry to be small, a priori.

\section{Genus: Examples and Mathematical Settings}

Genus is a quantitative measure of topology. It can characterize both 3D and 2D distributions. The two-dimensional genus can be represented as (Coles 1988, Mellot et al. 1989):
\\ \begin{equation}
 G \equiv (\mbox{number of isolated high-density regions}) - (\mbox{number of isolated low-density regions}).
\end{equation}
Where low- and high-density regions are selected with respect to a given density threshold. As a result, for a given 2D  intensity map, a curve corresponding to different thresholds emerge (see Fig. 1).

For instance, a uniform circle would have a genus of 0 (one contiguous region of high density, i.e. "island" and one contiguous region of low density, i.e. "hole") while a ring (a CD for example) would have a genus of -1 (one contiguous region of high density and two contiguous regions of low density). Two separate circles, one the other hand, correspond to the genus of 1 (one "hole" and two "islands"), three separate circles -- to the genus of 2 etc. Using the language of Richard Gott, one can say that genus can distinguish between the meatball and Swiss cheese topology.  

Furthermore, the genus can be represented mathematically as an integral using the Gauss-Bonnet theorem. In more specific terms for the 2D case we have \citep{M89,G90}:
\begin{equation} 
G(\nu) = \frac{1}{2\pi} \oint_{\mathcal{L}(\nu)} \! \kappa(x,y) \: dl
\end{equation}
\noindent where
\begin{equation*}
\mathcal{L}(\nu) \equiv \{ (x,y) \, | \, I(x,y) = \nu \}
\end{equation*}
\begin{equation*}
\kappa \equiv \frac{1}{r} \cdot \mbox{sign} (-\nabla I  \cdot \vec{n})
\end{equation*}
\noindent $I$ is an observed intensity, $r$ is the principal radius of curvature and the integral follows a set of contours of the surface at given $\nu$, and $\vec{n}$ is a normal vector, pointing outside of a contour. Much like in eq. (1), a contour enclosing a high-density region will give a positive contribution, while a contour enclosing a low-density one will give a negative contribution. Essentially, at a given threshold value $\nu$, the genus value is the difference between the number of regions with a density higher than $\nu$ and those with a density lower than $\nu$.

The threshold values in eq. (2) are selected so that they represent area fractions $f$. For a Gaussian field they are defined as \citep{H02}:
\begin{equation}
f = \dfrac{1}{\sqrt{2\pi}} \int_{\nu}^{\infty} e^{\frac{-x^{2}}{2}} dx
\end{equation}
\noindent Raising the threshold level $\nu$ from the mean value would cause the low-density regions to merge together, causing the genus to become positive, reach its maximum and begin to decrease to zero while positive regions begin to disappear with larger $\nu$. Similarly, lowering the threshold level $\nu$ from the mean value would cause the high-density regions to coalesce, resulting in a negative genus. 

More importantly however, the genus curve for a random Gaussian distribution is known (Coles 1988):
\begin{equation}
G(\nu)=\frac{1}{(2\pi)^{3/2}}\frac{\langle \kappa^{2}\rangle}{2} \nu e^{\frac{-\nu^2}{2}} = A \nu e^{\frac{-\nu^2}{2}}
\label{Gauss}
\end{equation}
This particular form of the genus curve characterizes a Gaussian random field, whatever its power spectrum is. This is an extremely important point of the genus analysis, because it allows us to separate topology effects from the ones caused by the power spectrum behavior. 

We expect, that the sign of the genus curve at the mean intensity level does describe the field topology. The positive genus will represent a clump-dominated field, while the negative one should mean the domination of holes. However it is more convenient to work with the zero of the genus curve $\nu_0$, because it can be naturally normalized to the field variance. In this case, for the intensity with the subtracted mean value the negative $\nu_0$ corresponds to the clumpy topology, and the positive $\nu_0$ indicates the ``swiss cheese'' one. 

\begin{figure}
\plottwo{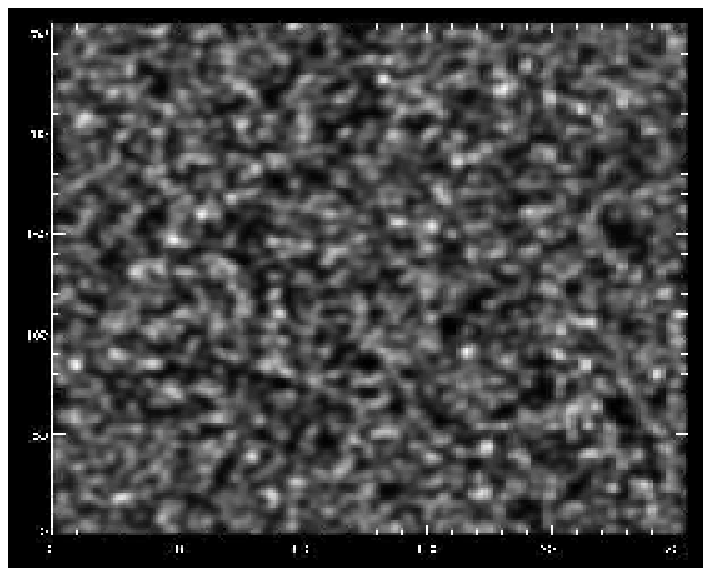}{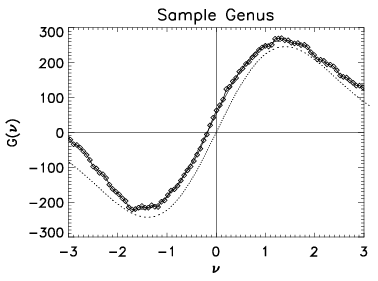}
\caption{Snapshot of the toy image created in IDL. The left-hand image displays the high-density clumps randomly distributed on a low-density background. The right-hand image displays the genus curve ($\lambda = 2$ pixels). It displays a typical clump topology: a negative shift of the genus curve zero. The dotted line corresponds to the Gaussian curve.}
\label{sample_clumps}
\end{figure}

An example of the genus curve is presented in Fig.~\ref{sample_clumps}. That is an example of clumpy topology, as $\nu_0<0$. An additional information on the distribution topology can be obtained by comparing the genus curve for the given distribution with the one for a Gaussian distribution but the same dispersion (see \ref{Gauss}). The points of maximum and minimum of the genus curve correspond to percolation of the distribution (see Colombi et al. 2003). The slower fall at large thresholds $\nu$ of the observed genus compared to the genus of the Gaussian distribution indicates that the islands are more discrete and pronounced than for the Gaussian distribution. 

As a Gaussian field always has a neutral topology, fitting of the Gaussian genus for estimation of $\nu_0$ can not be an optimal choice. As we do not rely on a particular field statistics, we need a robust method for estimation of $\nu_0$ for any form of a genus curve. In this paper we use fitting of a polynomial between the global extrema of a genus curve with additional condition of zero derivative at the ends of such interval. Practical calculations show, that the 5-th power polynomial is flexible enough to represent varying shape of non-Gaussian genera, and always has a single zero, which we interpret as an estimation of $\nu_0$. The higher polynomials tend to oscillate when applied to a noisy genus.

It can be shown on a simple example, that large-scale trends, even linear, can significantly distort a genus curve, changing its shape and making it noisier. Such large-scale gradients usually don't carry any topological information and should be removed before estimation of $\nu_0$. Here we have options like subtracting of a polynomial background, or Fourier filtering of low harmonics in the whole map or its particular region.

Another possible source of contamination is presence of non-abundant compact features, with the amplitude hign enough to affect the mean value. Such features should be either removed from the map, or weakened by reducing of the image contrast. Taking the median value instead of the mean one when calculating $\nu_0$ is also an option.

On the other hand, the presence of white Gaussian receiver noise would not change the topology, as it corresponds to completely symmetric genus. We can substantiate this statement as follows. Let us consider some small region near the intersection of the plane at the level nu and the map surface. If the map has the positive curvature in the direcion of gradient, adding of such noise shifts the genus count to the positive direction. If the curvature is negative, the shift is negative. On the other hand, the mean curvature at the mean level will be positive for clumps and negative for holes, which means that the genus count at this level will be shifted up for clumps and down for holes, i.e. it would not change its sign. This means, that the topology type cannot be changed by adding of such noise.

The analysis would be incomplete without estimation of variance of $\nu_0$. Finding of the variance differs our present study from the earlier ones (Lazarian 2004, Park \& Kim 2007). Following suggestions by a pioneer of the genus analysis Peter Coles, we generated for each map a set of images with randomly shifted phases of individual harmonics. This procedure causes the field to take Gaussian statistics, and therefore zero $\langle\nu_0\rangle$, but allows us to effectively estimate its variance. 

In more detail, we take FFT of the region being studied and assign the phase of each harmonic to a uniformely distributed in $[-\pi,\pi]$ random variable, keeping Hermitian conjugacy of the Fourier image. After inverse FFT we calculate the respective $\nu_0$. After repeating this procedure severel times we calculate the variance of $\nu_0$. In our case 10 times appeared to be enough for obtaining of a statistically relevant variance.

\section{Observation, Data Analysis and Results}

\begin{figure}
\plotone{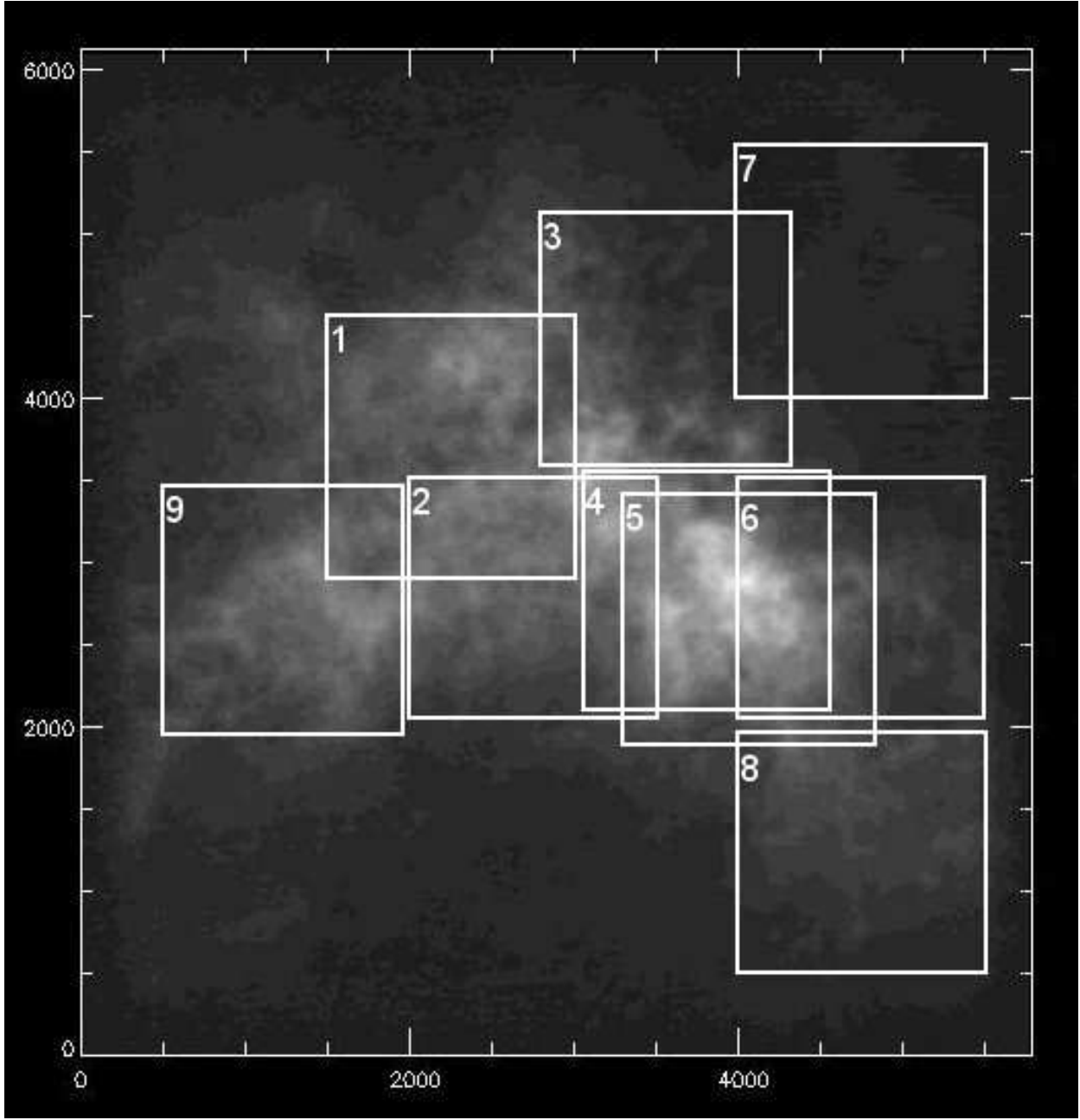}
\caption{HI greyscale HI column density image of the SMC from \cite{S99} with the studied regions. The image is a composite created from observations taken with the Parkes telescope and the ATCA. Black represents the minimum column density, while white represents the
maximum column density. The scale ranges from $0 \leq \mbox{intensity} \leq \mbox{1.03x}10^{22}$ atoms/cm$^{2}$. The spatial scale is given in pc.}
\label{SMC_ATCA}
\end{figure}

The HI column density image used in this study is a composite obtained with the Australia Telescope Compact Array (ATCA) and the Parkes Telescope in Australia (Fig. \ref{SMC_ATCA}). ATCA, a radio interferometer, was used to observe 320 overlapping regions containing the SMC. These data were combined with observations from the 64m Parkes radio telescope which observed a 4.5$^\circ$x4.5$^\circ$ region centered on RA 01$^{\mbox{h}}$01$^{\mbox{m}}$, Dec -72$^\circ$56' \citep{S99}. The data from these two telescopes were merged to create a complete image of the HI column density of the SMC, with a continuous sampling of spatial scales from 30 pc to 4.5 kpc. For more information on the merging process, see  Stanimirovic et al. (1999). The effective angular resolution of the combined column density image is 98$"$, implying  a spatial resolution of 30 pc at a distance of 60 kpc. 

The effective smoothing scale $\lambda$ is given here in terms of the effective HPBW and is always greater than 30 pc. For the 150x150 pixel region,  ($40 \mbox{ pc}\leq \lambda \leq 250 \mbox{ pc}$). Using a $\lambda$ larger than 15\% of the image length resulted in a genus curve that was useless - both the high-density and low-density regions coalesced together, resulting in too few regions for the genus statistic to analyze. The background was subtracted using a 5th order polynomial with subsequent filtering out of the first two Fourier harmonics.

\begin{figure}
\plotone{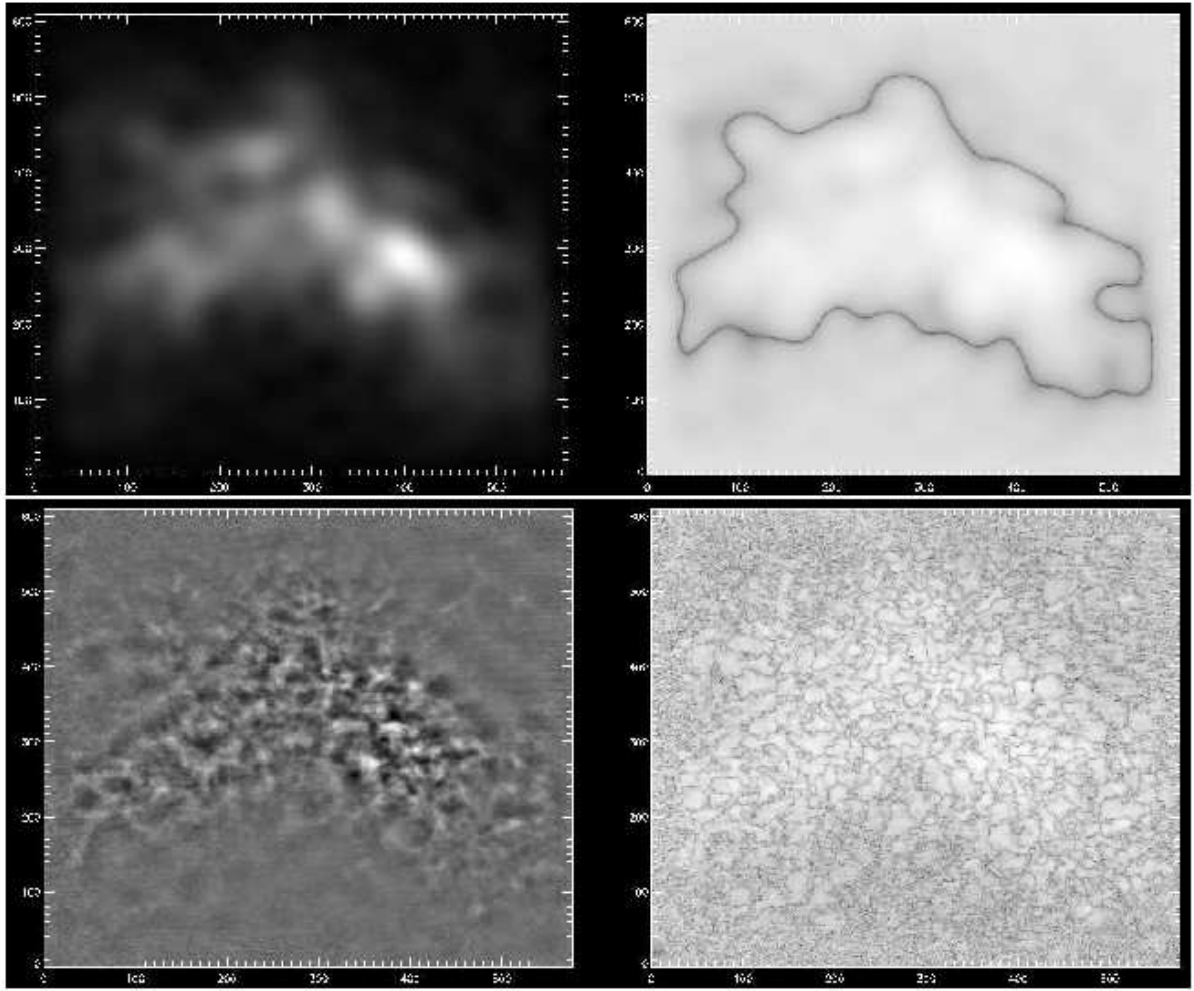}
\caption{Snapshots of the Fourier filtered SMC. The top image (left-hand column) is the SMC with the high frequencies excluded: the general shape of the SMC is apparent but no small-scale features are seen. The bottom image is the SMC with the low frequencies excluded: the numerous small-scale features of the SMC are readily apparent and much more visible. The counterpart of each image (right-hand column) is the absolute value of the natural log of the image, which results in the contour-like image on the right. }
\label{FFT}
\end{figure}

An offshoot of the Fast Fourier Transform (FFT) package in IDL was also utilized to study the HI column density image of the SMC. The SMC column density map was converted into Fourier space and non-informative frequencies were removed (see Fig.~(\ref{FFT})). An inverse Fourier transform was then applied to recover the information. This methodology allows us to decompose the image into different frequency ranges, thereby probing the SMC at different scales. By removing the low frequencies, it is possible to focus on the small scale structure of the SMC. Similarly, by removing the high frequencies we decrease the resolution and focus on the large scale structure of the SMC. This method works in conjunction with the smoothing scale methodology as
described above.

\begin{figure}
\plotone{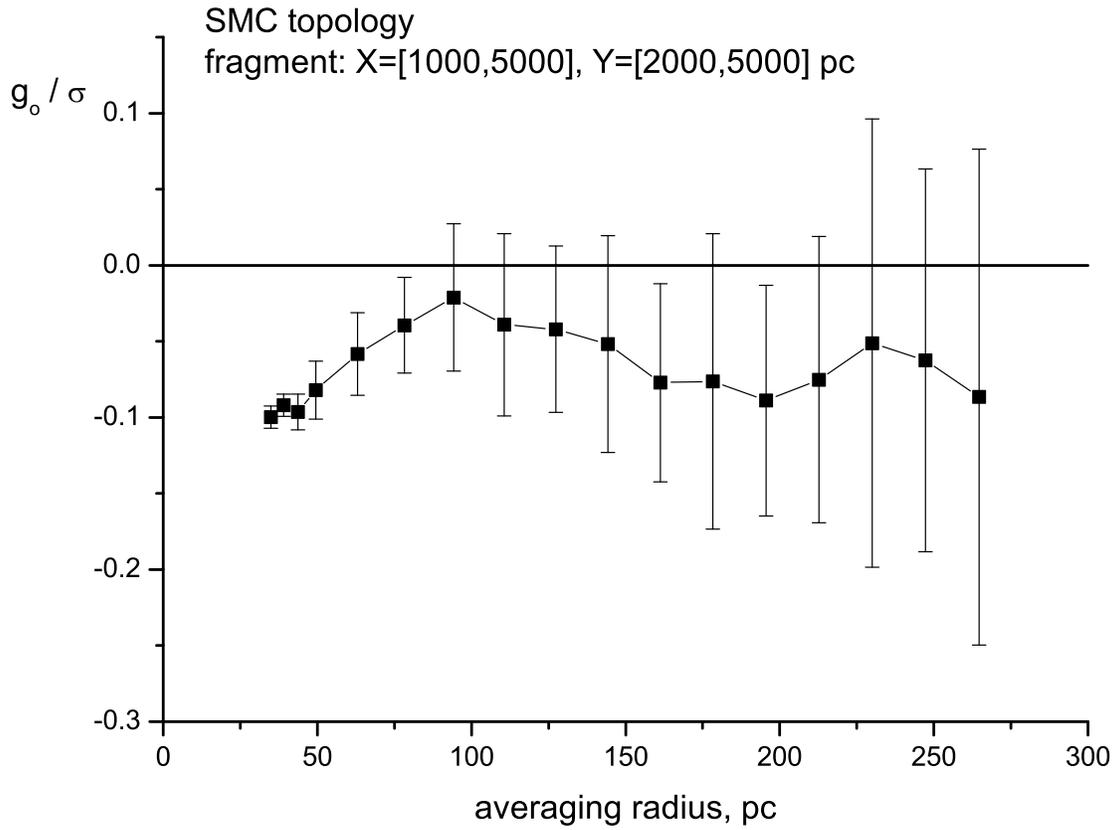}
\caption{Genus shift vs. smoothing radius for the entire SMC. 
The underlying astrophysics processes behind the genus shift are discussed in Sections 4 \& 5. }
\label{SMC_Entire}
\end{figure}

Figure \ref{SMC_Entire} shows the genus shift as a function of smoothing radius $\lambda$ for a 400x300 pixel region enclosing the majority of the SMC. At all smoothing scales ($40 \mbox{ pc}\leq\lambda\leq 270 \mbox{ pc}$), the SMC shows negative shift, being statistically significant at small ($40 \mbox{ pc}\leq\lambda\leq 80 \mbox{ pc}$) and medium ($\lambda \sim 170 \mbox{pc}$) scales. 

\begin{figure}
\begin{center}
\begin{tabular}{lll}
\inctab{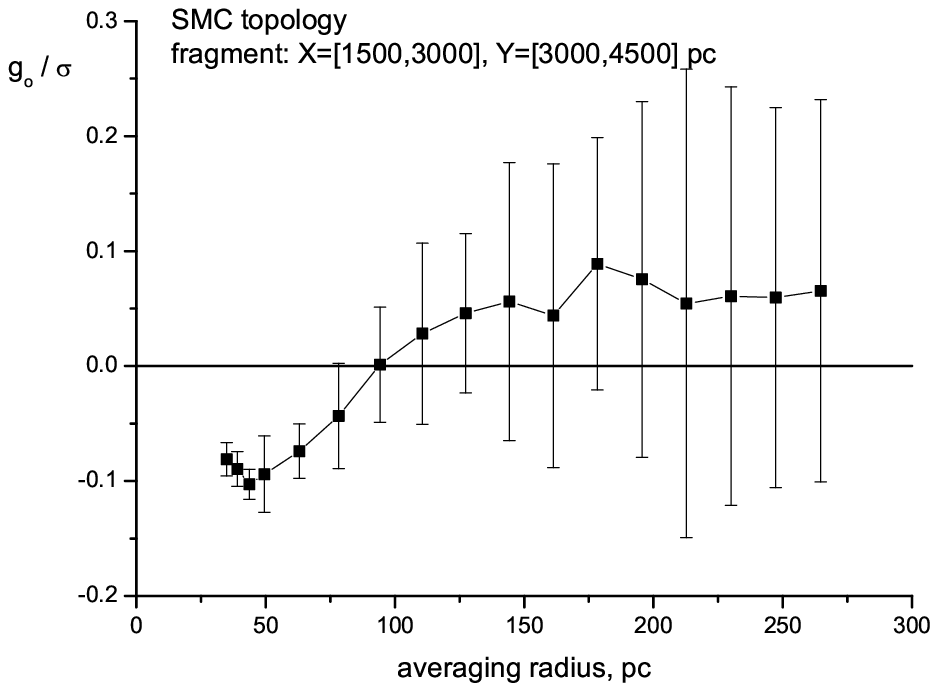} &
\inctab{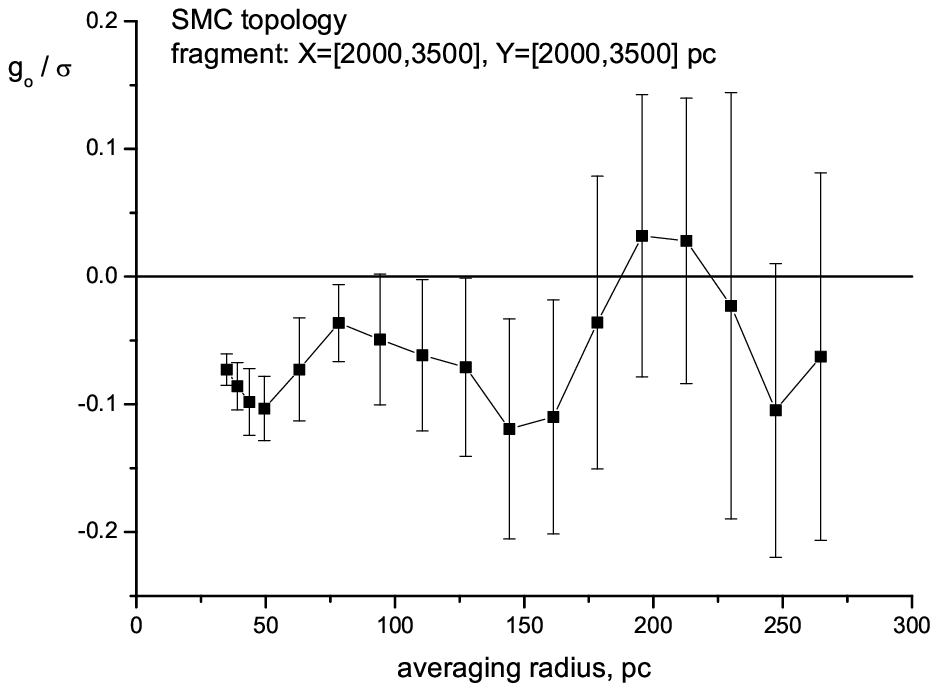} &
\inctab{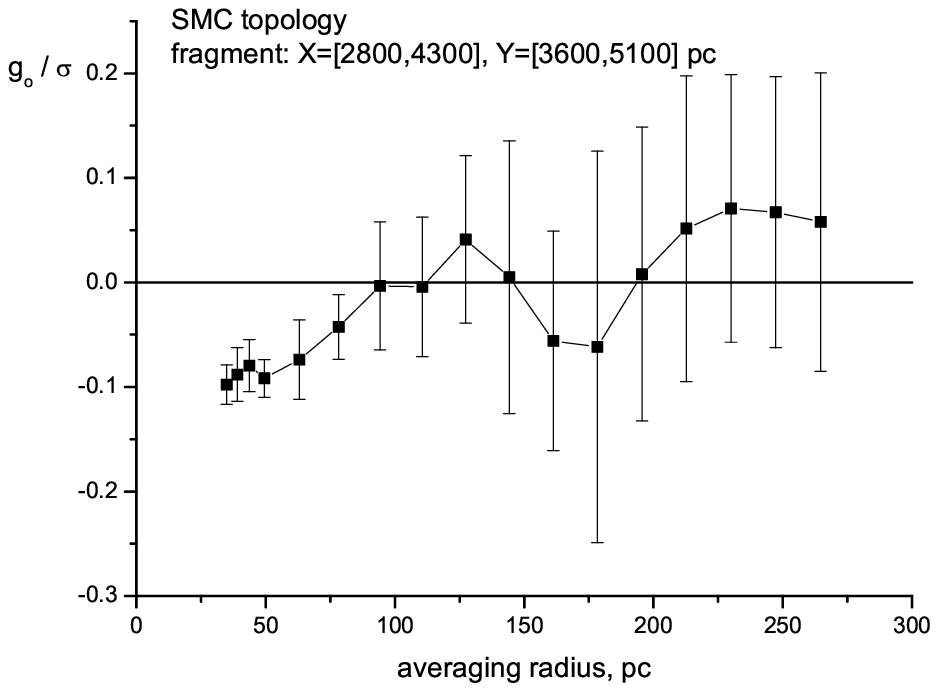} \\
\inctab{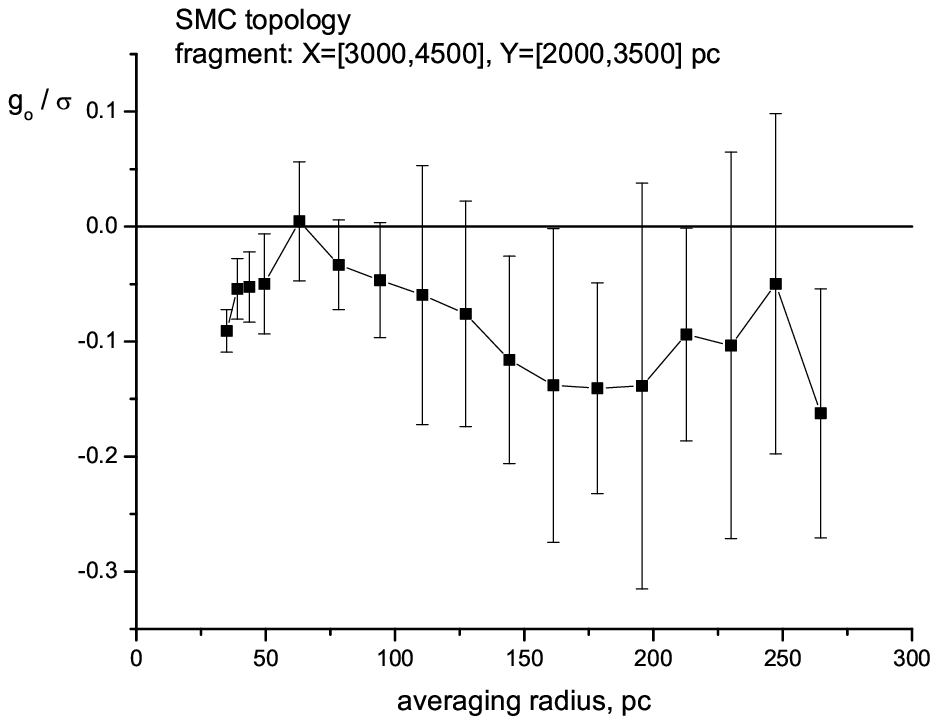} &
\inctab{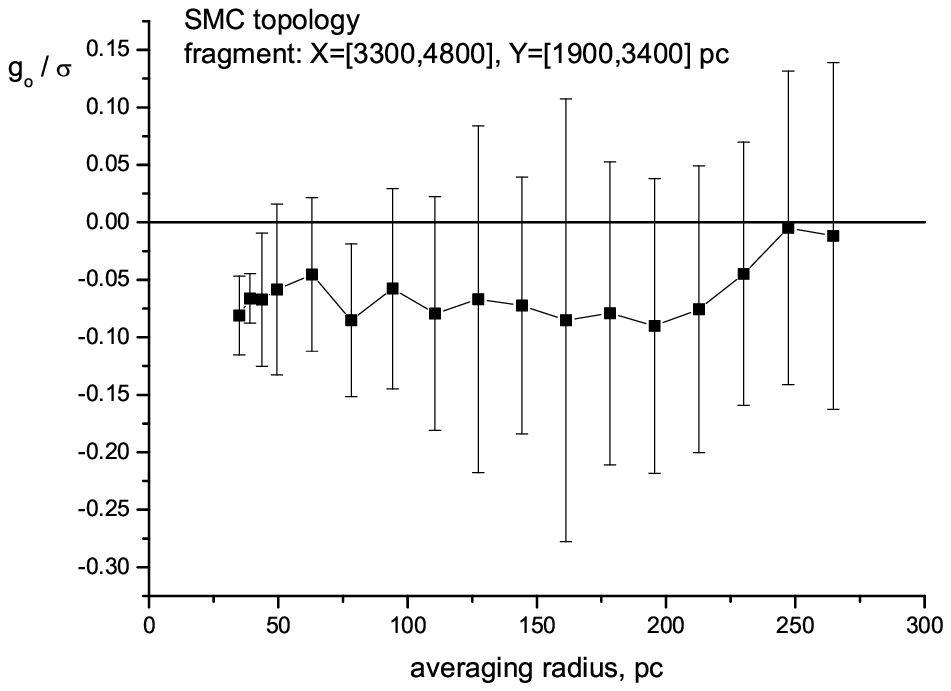} &
\inctab{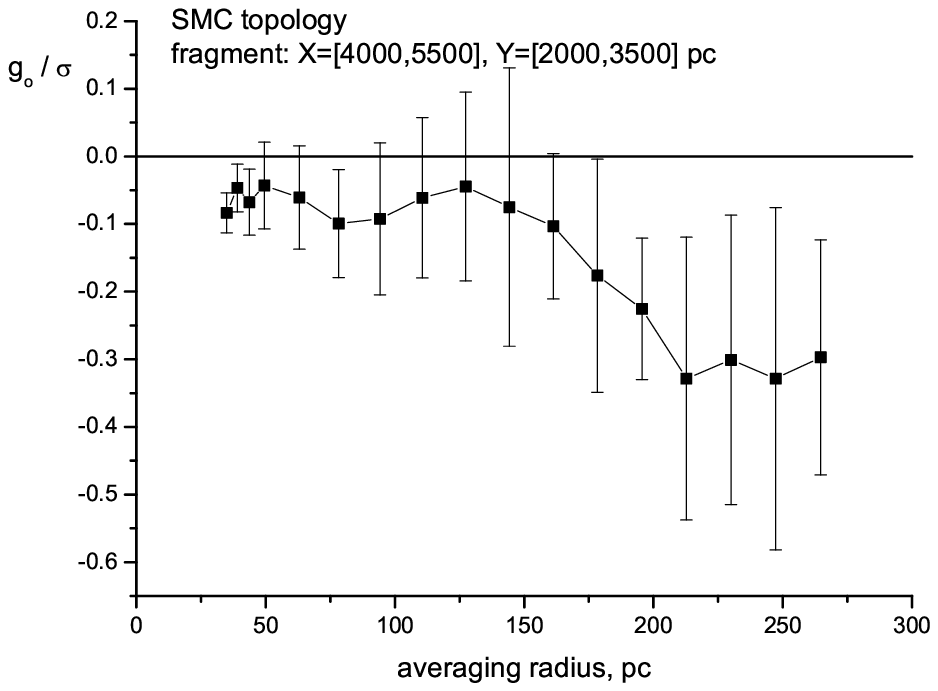} \\
\inctab{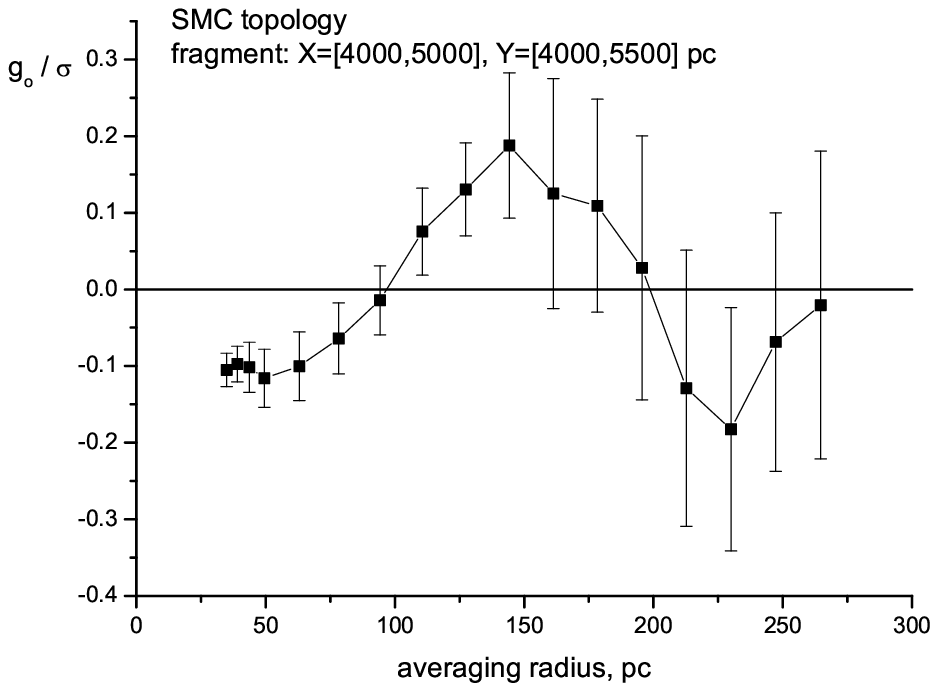} &
\inctab{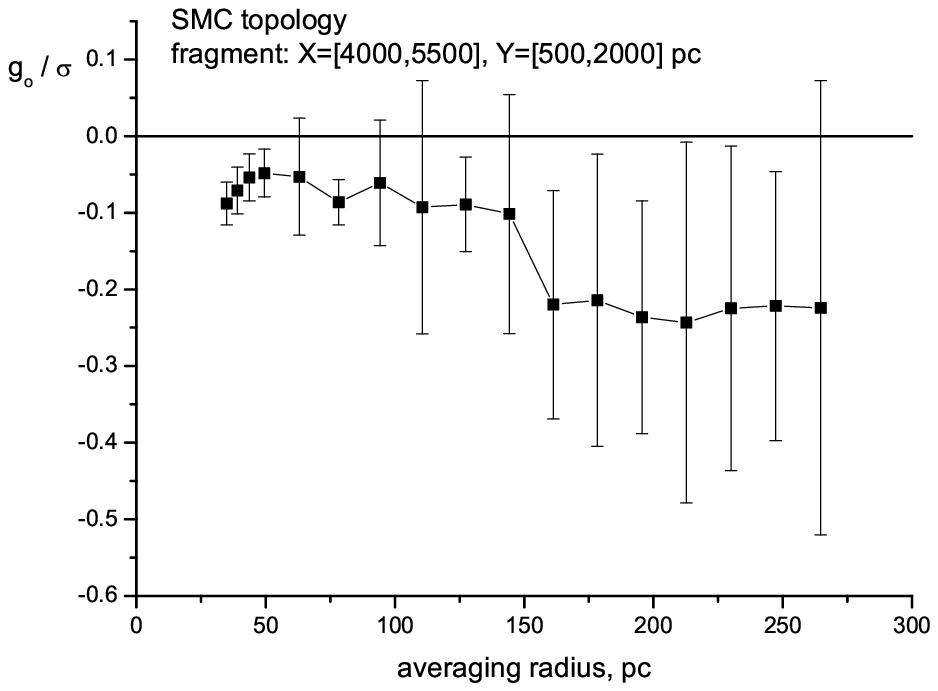} &
\inctab{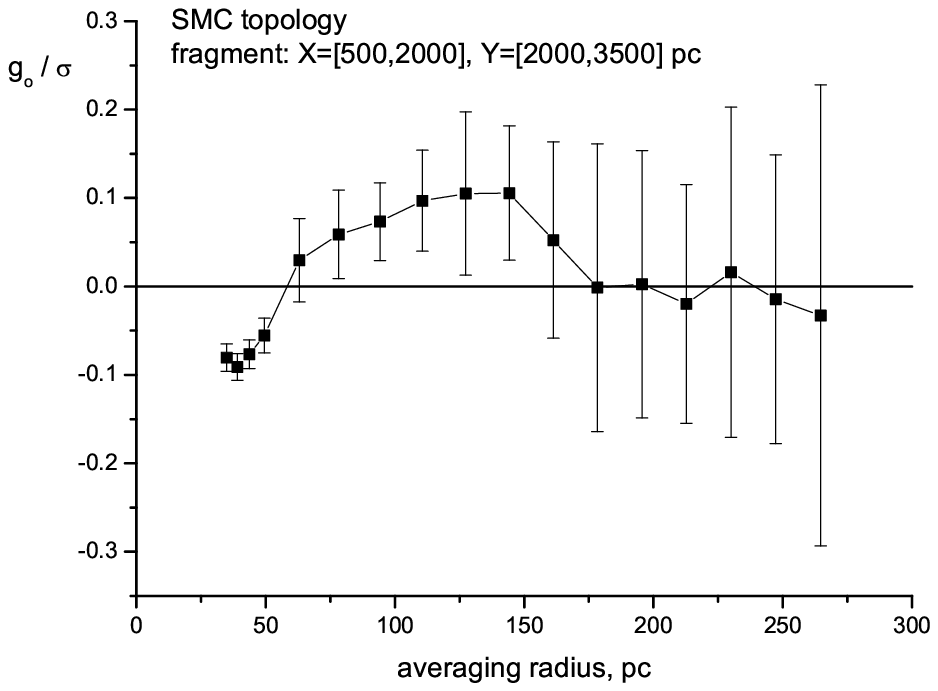}
\end{tabular}
\end{center}
\caption{Genus shift for the nine 150x150 pixel surveyed regions of the SMC. The underlying astrophysics processes behind the genus shift are discussed in Sections 4 \& 5. The effective resolution, accounting for the instrument HPBW, is used. The plots are ordered by region numbers (the first row corresponds to regions 1,2,3 etc.)}
\label{150x150_Genus}
\end{figure}

The results from the 150x150 pixel regions can be seen in Figure \ref{150x150_Genus} and Tab. \ref{tab:shifts}. At scales below  70-80 pc, every sampled region shows a genus curve with apparent negative $\nu_0$. Furthermore, at these scales each region exhibits asymmetry, though this varies from region to region. Five of the nine surveyed regions have a larger amplitude on the negative (low-density) side of the genus curve. We can infer from this asymmetry that the low-density holes are more isolated than
expected while the high-density clumps are more contiguous than expected. Combined with the negative genus shift, these two statistics show that there are merged high-density clumps surrounded by isolated low-density holes. The negative shift can be attributed to the numerous small clumps of gas which compose the ISM of the SMC. At the small scales at which the genus is probed, the numerous shells dispersed throughout the SMC are not seen. 

The genus shift curves for the individual regions diverge as the smoothing scale increases. At scales of 120 to 150 pc, two of 150x150 regions show a positive genus, implying ``swiss-cheese'' topology. We can conclude from the rising genus shift that the small clumps are merging together while the holes and SGSs are coming into focus. Four regions have a negative shift at large scales (150-250 pc), and one of them shows mixed behavior (positive genus at 120-150 pc and negative genus at 220-270 pc). The error bars obtained by randomly applying phases to Fourier modes as described in \S 2 show that for some scales and for some parts of the SMC our results are more reliable than for others.

\begin{figure}
\begin{center}
\begin{tabular}{l}
\inctabp{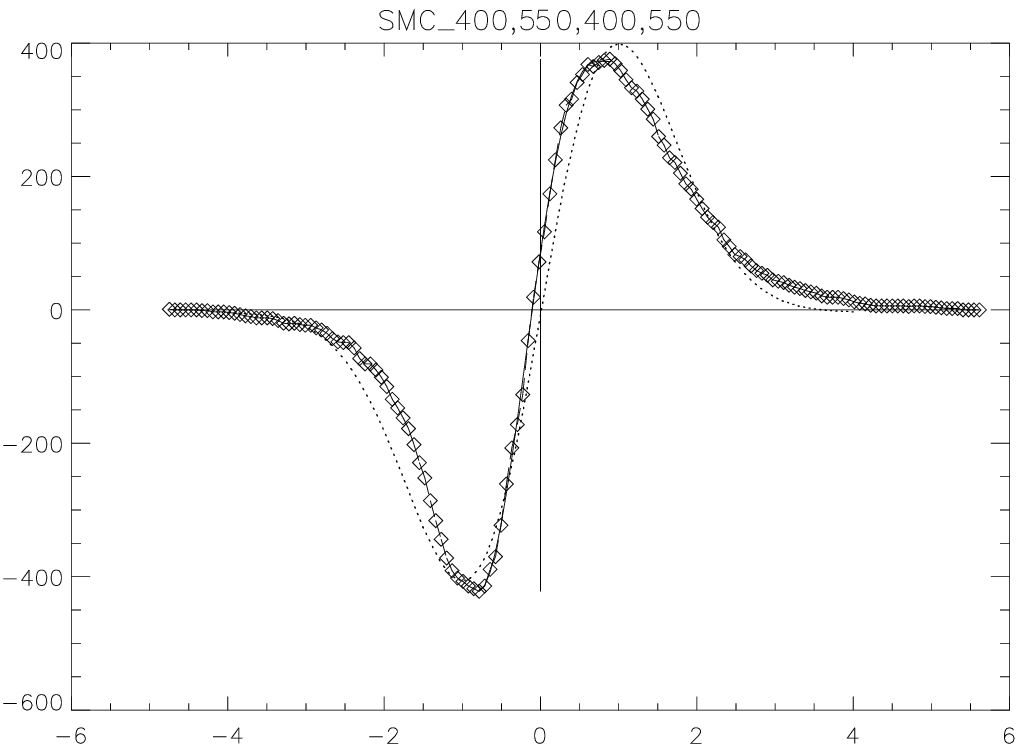} \\
\inctabp{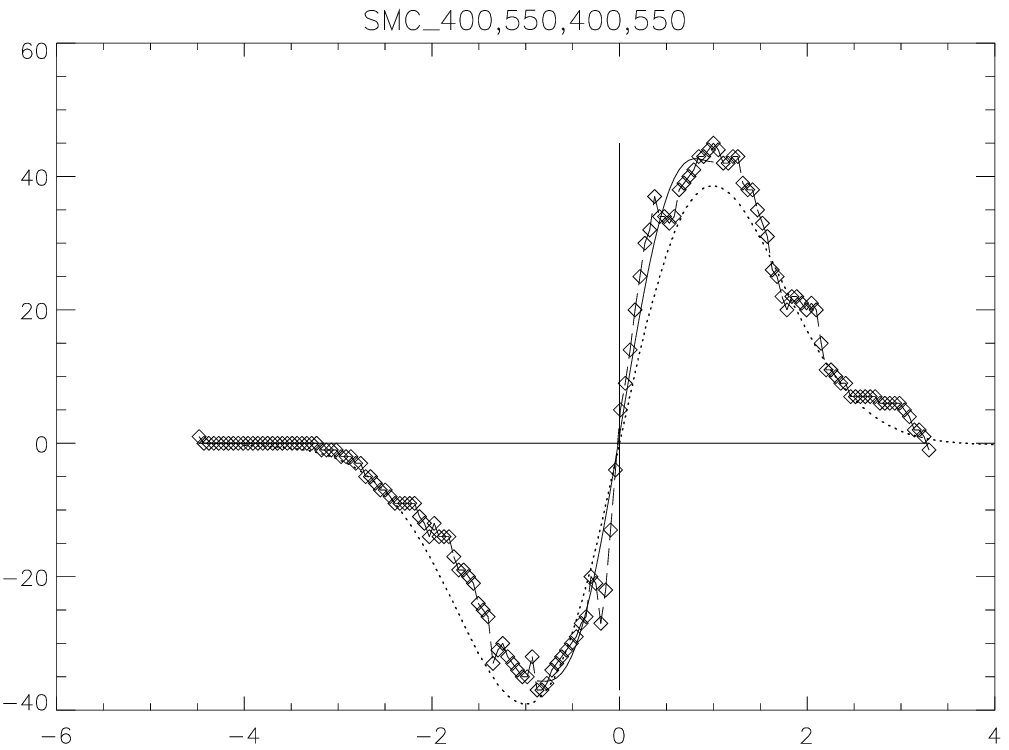} \\
\inctabp{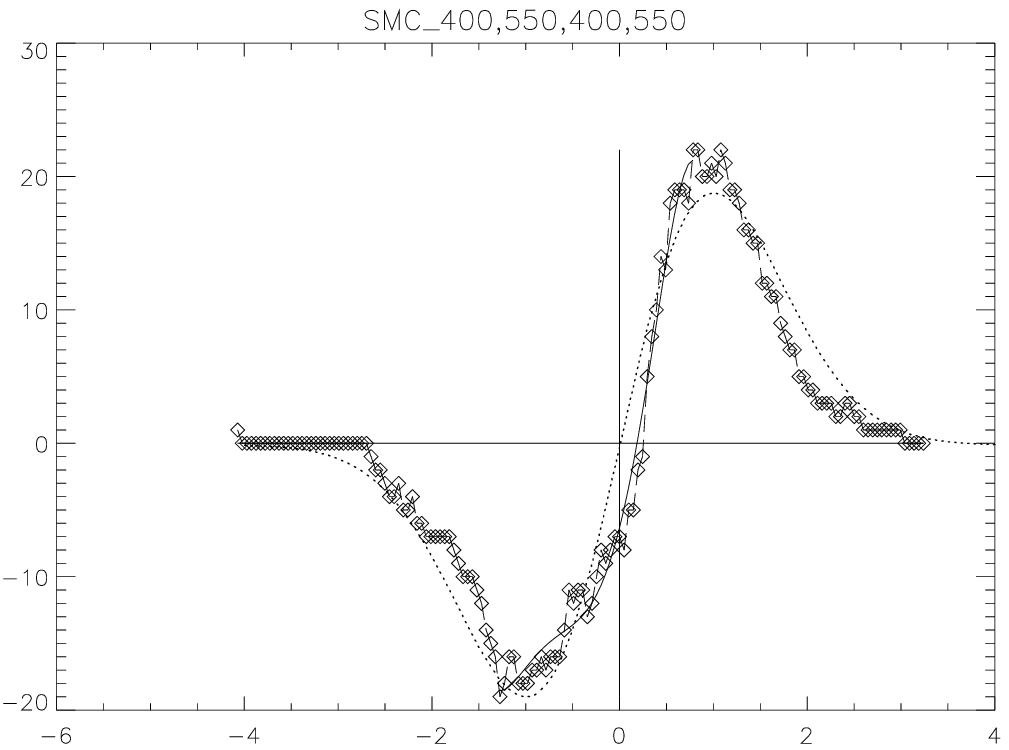} \\
\end{tabular}
\end{center}
\caption{Genus curves for the region 7 with mixed topological behavior for different smoothing scales. The graphs correspond to $\lambda$ equal to 35 pc, 93 pc and 144 pc, displaying ``meatball'', neutral and ``swiss cheese'' topologies. Dashed curves correspond to the fitted polynomial. Dotted lines correspond to the genuses for the Gaussian distributions.}
\label{Genera}
\end{figure}

\begin{figure}
\plotone{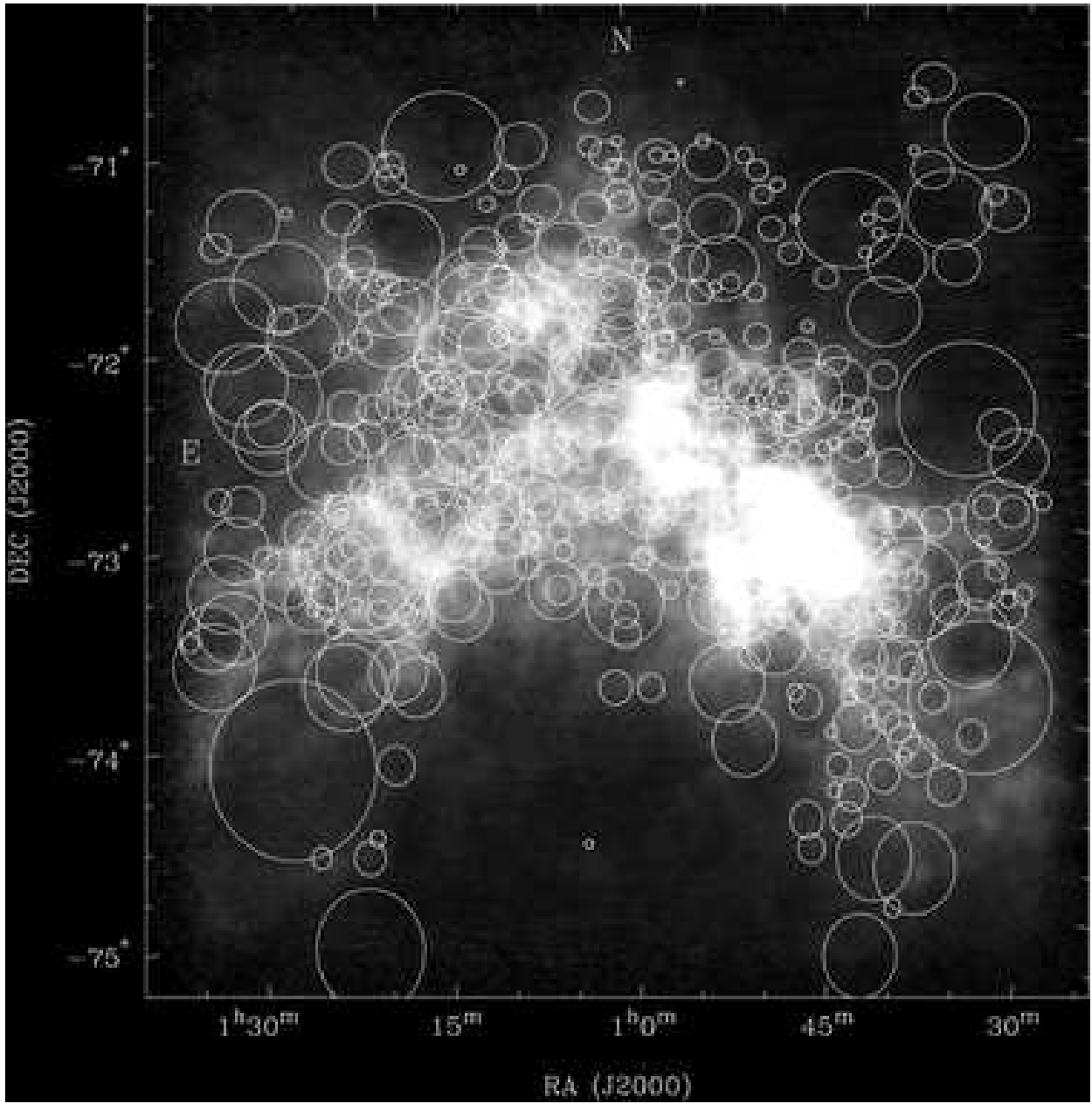}
\caption{S
hells, detected in SMC, plotted on the HI intensity map}
\label{Shells}
\end{figure}

The example of genus curves for an individual region with mixed topological 
behavior for different $\lambda$ is shown on Fig. \ref{Genera}.

\section{Discussion}

The genus shift estimated in the previous section can give us insight into the underlying physical processes of the SMC. Referring to the genus shift of the entire SMC (see Figure \ref{SMC_Entire}), it is readily apparent that the shift varies according to the smoothing scale $\lambda$. At the smallest scales studied, the genus shift has a negative value, implying a strong clump topology. We can infer that the clumps are caused by clouds of HI gas, as well as the  numerous knots and filaments that compose the SMC. At medium scales (120-200 pc), the genus shift takes on a neutral or slight positive value. This increase in the
genus shift can be connected to the abundance of shells that compromise the SMC, as the $\sim500$ shells which are interspersed throughout the SMC have a mean radius of $\sim100$ pc \citep{SS07}. This is consistant with our results, as the largest positive values of the genus shift occur between 120 and 150 pc.  At the largest scales studied ( $>$ 170 pc), the genus shift takes on a slight negative value;  this is due to the prominent `wing' and `bar' features \citep{S99}, indicating that at even the largest scales, (large) clumps dominate the topology.

From our results, we reached the conclusion that the SMC tends to exhibit a clump topology. The dominance of the ``meat ball'' topology is expected in the case of supersonic turbulence (Kowal et al. 2007\footnote{The adiabatic approximation, used in the latter work, may not be valid for all the studied here scales, but we still can use its results as a guidance.}).  Several of the surveyed regions display characteristics similar to those of supersonic (low $\beta$ case) -- an asymmetrical genus curve with a tail that extends into the high-density portion of the plot (see \cite{KLB07}, figure 18 for more information). However, other possible reasons for a clump topology could be cooling instability and self-gravity of the HI gas.

However, our results differ from the results in Kim \& Park (2007) who found that the HI distribution in the LMC shows mainly hole topology at intermediate scales, despite the fact that fewer shells are present in the LMC (124 as compared to $\sim500$ in the SMC). There are several important factors that could contribute to this difference. Firstly, Kim \& Park (2007) performed genus analysis on the HI peak brightness temperature image, while our work used the HI column density distribution. Peak brightness images emphasize more small-scale and shell structure, while the column density distribution emphasizes density distribution washing out small-scale fluctuations that are caused both by density and velocity fluctuations. 
Secondly, while the LMC has an almost face-on HI disk, the SMC has a larger inclination and a non-disk-like morphology. In fact, several authors have claimed a large line-of-sight depth of the SMC (see Stanimirovic, Staveley-Smith, Jones 2004 for details). Therefore, any line-of-sight through the SMC integrates over a longer physical depth. It would be interesting to apply our procedure on the LMC HI column density image for a more direct comparison.

Interestingly, for several regions where a hole topology is detected, the estimated shell size is comparable to the one from Kim \& Park (2007). A possible reason for a hole topology in regions 7 and 9 could be related to the shells
created by supernovae explosions and stellar winds. It is interesting to note that those two regions are both off the main `bar' of the SMC and most likely correspond to areas of low line-of-sight depth. Nigra et al. (2008, in preparation) study the Eastern Wing region (our region 9) and find a small line-of-sight thickness. Our region number 7 has been identified in Stanimirovic et al. (1999) and \cite{H05} as containing several `orphan' shells with high luminosity. \cite{H05} hypothesize that these shells may be associated with an ancient chimney.

We stress, that it is important to gauge the statistical techniques against numerical simulations. \cite{KLB07} recently undertook a extensive investigation of density statistics in MHD turbulence, which included different statistical tools, including genus. They concluded that the genus statistic \textit{is} is sensitive to the sonic Mach number \textit{$M_{s}$}. In the case where the magnetic pressure dominates, i.e. the high-$\beta$ case, and subsonic Mach number (e.g. $M_{s}$ $\approx$ 0.3) the genus curve  is highly symmetrical. For the low-$\beta$ cases and supersonic (e.g. $M_{s}$ $\approx$ 2.1, 6.5), the curve stretches into the positive (high-density) side and becomes increasingly non-symmetrical as $M_{s}$ increases. The end result is that it is possible to obtain $M_{s}$ from the plot of genus curve: the Mach number is directly related to the length of the high-density tail. Furthermore, the genus statistic also gives topological information. For values of low $M_{s}$, the genus curve is symmetrical,
implying that there are equal numbers of high-density clumps and low-density holes. As $M_{s}$ increases, turbulence creates more high-density structures, which causes the tail of the genus curve to extend towards the high-density side. This corresponds well to the results of studying fractal dimension of density while varying the density threshold in Kowal \& Lazarian (2007). 

In our analysis we assumed that the HI gas is optically thin. This is an important assumption, as the analysis of the effects of absorptions in turbulent gas in Lazarian \& Pogosyan (2004) suggests that absorption introduces a critical spatial scale for plane-of-sky statistics, with the fluctuations larger than this critical scale being strongly affected by absorption effects. This means that the genus the way we use it above is not applicable to $^{12}$CO data, but should be applicable to C$^{18}$O data, with
$13$CO data is in limbo, as this isotope is frequently thick. Potentially, the topology of gas at scales less than the critical scale is also of interest. However, Lazarian \& Pogosyan (2004) study suggests that fluctuations of the integrated self-absorbing spectral line may be dominated by velocity caustics, provided that the spectrum of density is sufficiently steep, e.g. $E(k)\sim k^{-\aleph}$, $\aleph>1$. The Kolmogorov spectrum corresponds to $E(k)\sim k^{-5/3}$ and is steep according to the aforementioned definition. The topology studies are, as we discussed above, are different from the studies of spectrum.  However, we believe that the criteria for density fluctuations being observable is the same. Therefore, we expect that for $\aleph<1$ one can analyze genus for the scales less than the critical one. Note, that in terms of the present study, the constancy of the spectral indexes observed in Stanimirovic \& Lazarian (1999) provides an additional evidence of HI not being substantially affected by absorption. 

\section{Summary} In the paper above, we have analyzed the HI column density map of the SMC in an attempt to elucidate its topological features. A brief summary of our results is as follows:
\begin{itemize}
\item We have extended the genus analysis for column densities of diffuse gas via presenting a new procedure for estimating of topology indicator $\nu_0$ and its variance.
\item At small scale of smoothing ($35 \mbox{ pc}\leq\lambda\leq 120 \mbox{pc}$) the SMC exhibits a negative shift, indicating a clump or "meatball" topology. We conjecture that this is due to the numerous clumps of gas created by supersonic turbulence. We know from numerical simulations that numerous high contrast clumps are produced by such a turbulence. 
\item As the smoothing scale increases ($120 \mbox{ pc}\leq\lambda\leq 150 \mbox{ pc}$), the shift of the genus curve becomes less negative, trending towards a slight positive shift. This can be attributed to the averaging of small clumps, while larger shell and SGS structures throughout the SMC are less affected by smoothing. At these medium scales, the smaller gas clumps are less important, while the shells come into focus. These shells are potentially a result of stellar winds and SNe from OB associations.
\item For larger regions with scales $\geq$ 100pc the genus curve becomes noisier, however in four cases the correspondent negative shift may indicate that most of the shells have sizes less than the smoothing scale.
\item The nine 150x150 pixel regions of the SMC exhibit slightly different trends. Although they all possess a clump topology at small scales, the curves at larger scales are rather different. Some trend towards a hole topology at larger $\lambda$ while others exhibit no positive genus shift. A possible reason for hole topology in regions 7 and 9 could be related to the shells
created by supernovae explosions and stellar winds. We hope that in future the regions with particular topology can be identified with regions of physically distinct behavior. We may infer that the SMC is somewhat heterogeneous from region to region. 
\item Genus analysis is an effective complementary tool in the study of turbulence. The power spectrum contains information on velocity fluctuations but does not possess topological information. Combined with genus statistics, both the velocity statistics and topological information can be obtained for a selected object.
\end{itemize}

We thank Peter Coles for his important input on the genus analysis and Dmitry Pogosyan for his suggestions. J.G. would like to acknowledge the help he received from A.C., A.L and S.S. during the summer REU program. J.G. is supported by the NSF Research Experience for Undergraduates (REU) program. A.L. acknowledges NSF grant AST 0307869 and the Center for Magnetic Self-Organization in Astrophysical and Laboratory Plasmas. We also thank the anonymous referee for a number of valuable points.

\begin{deluxetable}{lllll}
\tablecaption{Statistically significant genus shifts \label{tab:shifts}}
\tablehead{
  \colhead{region id} & 
  \colhead{$X$, pc} & 
  \colhead{$Y$, pc} & 
  \colhead{scales, pc} & 
  \colhead{$\nu_0/\sigma$}
}
\startdata
1  & 1500-3000 & 3000-4500 & 35-70   & $-0.10 \pm 0.02$ \\
2  & 2000-3500 & 2000-3500 & 35-70   & $-0.09 \pm 0.02$ \\
   &           &           & 150-170 & $-0.11 \pm 0.09$ \\
3  & 2800-4300 & 3600-5100 & 35-70   & $-0.09 \pm 0.03$ \\
4  & 3000-4500 & 2000-3500 & 35-50   & $-0.06 \pm 0.03$ \\
5  & 3300-4800 & 1900-3400 & 35-40   & $-0.07 \pm 0.04$ \\
6  & 4000-5500 & 2000-3500 & 35-40   & $-0.07 \pm 0.04$ \\
   &           &           & 200-270 & $-0.30 \pm 0.19$ \\
7  & 4000-5500 & 4000-5500 & 35-100  & $-0.09 \pm 0.04$ \\
   &           &           & 120-150 & $ 0.13 \pm 0.07$ \\
   &           &           & 220-230 & $-0.18 \pm 0.16$ \\
8  & 4000-5500 & 500-2000  & 35-50   & $-0.07 \pm 0.03$ \\
   &           &           & 170-250 & $-0.23 \pm 0.19$ \\
9  &  500-2000 & 2000-3500 & 35-50   & $-0.08 \pm 0.02$ \\
   &           &           & 120-150 & $ 0.09 \pm 0.06$ \\
\enddata
\tablecomments{(The confidence intervals are given here for confidence probability 0.67)}
\end{deluxetable}

\end{document}